# Experimental Realisation of Dual Periodicity Moiré Superlattice in a MoSe$_2$/WSe$_2$ Heterobilayer

*Liam P. McDonnell[1], Jacob J. S. Viner[1], Pasqual Rivera[2], Antonio Cammarata[3], Huseyin S. Sen[3], Xiaodong Xu[2], David C. Smith[1*]*

[1] School of Physics and Astronomy, University of Southampton, Southampton SO17 1BJ, United Kingdom.

[2] Department of Physics, University of Washington, Seattle, WA, USA.

[3] Department of Control Engineering, Faculty of Electrical Engineering, Czech Technical University in Prague, Technicka 2, 16627 Prague 6, Czech Republic.

Abstract (~147 words)

Moiré structures in van der Waals heterostructures lead to emergent phenomena including superconductivity in twisted bilayer graphene[1,2] and optically accessible strongly-correlated electron states in transition metal dichalcogenide heterobilayers[3–5]. Dual periodicity moiré structures (DPMS) formed in layered structures with more than two layers have been shown to lead to ferromagnetism [6] and multiple secondary Dirac points in TBG [7]. Whilst in principle it is possible to obtain DPMS in bilayers there has not been clear experimental evidence of this yet. In this paper we present signatures of DPMS in a twisted MoSe$_2$/WSe$_2$ bilayer revealed by resonance Raman spectroscopy. We observed zone-folded acoustic and optical phonon modes with a wavevector twice of the moiré wavevector, evidence of a dual periodicity moiré heterostructure. These results simultaneously open up opportunities for new emergent phenomena and an optical method for characterising DPMS in a wide range of van der Waals heterostructures.

**Introduction**

Moiré effects significantly modify band structures in van der Waals heterostructures leading to a wide range of new physical phenomena. The formation of flat electron bands in twisted bilayer graphene leads to both superconductivity[1] and emergent ferromagnetism[6], and in transition metal dichalcogenide (TMD) bilayers to optically accessible strongly-correlated electron states[3–5]. In the TMDs intralayer exciton moiré sub-bands [8] with sub-band separations of the order of 50 meV have been experimentally demonstrated and interlayer excitons in the TMDs also show significant modification of their band structure due to moiré structuring[9].

Whilst most of the moiré related phenomena have been explored in structures with a moiré potential with a single dominant periodicity, it is possible to produce structures with more than two layers with potentials with multiple periodicities [10]. This includes structures with periodicities with quite different scales, e.g. 2-10 nm and 10-100 nm, in a single structure. The effects of the double periodicity on the electrical transport [10,11] of hBN-graphene-hBN structures has been demonstrated [7] however there are only limited experimental results on these structures; at least in part because of the complexity of aligning multiple layers.

Another way to obtain multiple periodicity moiré structures (MPMS), which only requires two layers, is via superlattice effects. These occur for specific twist angles when the atoms go in and out of near alignment multiple times before achieving perfect realignment (Fig 1). The moiré potentials produced by such superlattice effects will have a strong modulation whose periodicity is given by the distance over which the atoms first come back into near realignment (moiré length scale). The magnitude of this modulation will be modulated with a periodicity associated with distance required to come back into perfect alignment (superlattice length scale). Whilst these have been proposed theoretically[12], until now these superlattice effects have not been experimentally verified. In this paper we set out the first experimental evidence for superlattice effects in a TMD heterostructure and show that resonant Raman scattering is the ideal tool for characterising the moiré periodicities and exploring the effect of these more complex moiré potentials on intralayer excitons.

**Main**

The results presented here are part of a study of high quality hBN encapsulated $MoSe_2$ and $WSe_2$ monolayers [14] and $MoSe_2$/$WSe_2$ heterobilayers [13], with twist angles 2, 6, 20, 57 and 60 °, using resonance Raman at 4 K and in the low excitation intensity regime (<100 µW). The Raman scattering intensity has been calibrated using absolute scattering rates for the silicon peak at 520 cm$^{-1}$ and corrected for Fabry-Perot effects [15,16]. Background luminescence has been reduced by subtracting cross-linearly-polarised spectra from co-linearly-polarised spectra and if necessary, by the fitting and subtraction of a slowly varying function (see SI Fig S1).

In Figs. 2 and 3 we present the key Raman peaks relevant to this paper. In Figs. 2 a and b we present spectra obtained by resonantly exciting the $MoSe_2$ A1s and B1s excitons. In addition to the standard $MoSe_2$ $A_1'$ peak (~240 cm$^{-1}$), we observe a new peak, at 236.8 cm$^{-1}$ in the 2 ° sample and 234.8 cm$^{-1}$ in the 57 ° sample. These new peaks are not observed in monolayer $MoSe_2$ or any of the other heterobilayer samples (see SI Figs S2-S4). We attribute these peaks to moiré zone-folded satellites of the main $A_1'$ peak. No equivalent zone folded peaks are observed near the $WSe_2$ $A_1'$ peak (~249 cm$^{-1}$) even when resonant with the $WSe_2$ excitons, as expected due to the relatively flat dispersive nature of this phonon in this material (see SI Fig S9). In Fig 3 we present low frequency Raman shift spectra measured on the 57 ° twist sample. In these we observe a series of peaks when resonant with the A1s and B1s intralayer excitons associated with the two parent materials; we attribute these to moiré zone-folded longitudinal acoustic (LA) and transverse acoustic (TA) phonons. No equivalent peaks are observed in any of the monolayer or other heterostructure samples (see SI Fig S2-S4). In the 57 ° sample, with resonant excitation of the $MoSe_2$ A1s and B1s and $WSe_2$ A1s excitons (see Fig 3), the low frequency Raman spectra show two dominant peaks with satellite peaks at higher and lower shifts whose magnitude/presence depends on the excitation energy. The main peaks are strong; at the resonant excitation of $MoSe_2$ A1s exciton, their count rate is 50% bigger than that of the normally dominant $A_1'$ mode. Whilst the spectra resonant with the $WSe_2$ and $MoSe_2$ excitons are similar, the peaks observed at the $WSe_2$ resonant excitation have shifts approximately 9 % lower.

One of the reasons we attribute the new Raman peaks to moiré zone-folded phonons is there similarity to peaks observed in twisted MoS$_2$ homobilayers [12,17]. Whilst Liao et al. only observe a zone folded A$_1'$ phonon [17], Lin et al. report a more comprehensive set of moiré peaks with one for each of the three acoustic phonon branches along with moiré satellites of four of the optic phonons including the A$_1'$ [12]. The shifts of the peaks observed by Lin et al agree with those predicted using the monolayer phonon dispersion relation and the expected moiré wavevector. Whilst there is a clear similarity between the zone-folded moiré phonons reported in homobilayers and our samples, there are also several distinct differences. Firstly, from our resonance Raman experiments we observe a different set of low frequency Raman peaks when exciting either the MoSe$_2$ or WSe$_2$ intralayer excitons. Another key difference between our spectra and those reported by Lin et al. are the number of low frequency Raman peaks observed. Lin et al observe 3 peaks, whereas we observe up to 5 at the MoSe$_2$ resonances and 4 at the WSe$_2$ resonances.

To test the hypothesis that the additional peaks observed in the 2 and 57° samples can be explained by zone-folding of monolayer-like phonons, the spectra shown in Figs 2 and 3 were fitted to obtain the centre shift of the various peaks. The agreement of the shifts of the peaks observed at the various resonances (see SI) allow us to identify a single set of peaks associated with each material which are given in Table 1. Using DFT calculated phonon dispersion relations for monolayer WSe$_2$ we can determine the two wavevectors required to explain the shifts of the peaks at 52.9 and 57.8 cm$^{-1}$, assuming they are associated with the highest frequency (LA) acoustic phonon branch. We propose that all the other twist angle dependent peaks can be explained in terms of a central wavevector **k$_c$** and two sidebands with wavevectors **k$_{s+}$** = **k$_c$** + **Δk** and **k$_{s-}$** = **k$_c$** - **Δk**, where **k$_c$** and **k$_{s+}$** are the wavevectors associated with the 52.9 cm$^{-1}$ 57.8 cm$^{-1}$ peaks respectively. Using these wavevectors we then predict the frequencies for moiré zone-folded TA, LA and A$_1'$ phonons in both parent monolayers; a pictorial representation of this model is presented in the SI (see Fig S5). As can be seen in Table 1, there is remarkable agreement between the predicted and observed peak shifts to within ~ 1.0 cm$^{-1}$. Whilst the MoSe$_2$ TA peaks are all predicted systematically low, this can easily be explained by small errors in the predictions of the acoustic phonon dispersion relations. The agreement with the triplet of peaks near 35 cm$^{-1}$ in MoSe$_2$ is clear evidence that the two outer peaks are sidebands, rather than the triplet being due to three unconnected periodicities. A superlattice structure is the obvious explanation for a central peak with sidebands in which case the sideband separation gives the length scale associated with the superlattice which in this case would be 25.0 nm.

Whilst we do not observe zone folded peaks associated with all of the monolayer phonons this is not unexpected as the zone-folded optical phonons occur at shifts which overlap with multi-phonon peaks [14,18,19] making them hard to observe (see SI FigS7 & S8), whilst several of the acoustic modes would occur at too low a shift to be observed using our Raman system. However, we do not observe the ZA-derived layer breathing mode (LBM) branch phonon expected in bilayers. Based upon DFT calculations (see SI) and previous experimental measurements of the LBM Raman shift [20–22], we predict that the LBM should be observed between 20-30 cm$^{-1}$ but do not observe a peak at this shift when resonant with any of the intralayer excitons. An obvious explanation for the failure to observe this peak is that, unlike the acoustic modes we do observe, the layer breathing mode does not distort the lattices of the monolayers but instead just changes their separation. It is thus possible that it only very weakly couples to the intralayer excitons. Layer-breathing and shear modes have been observed but only at higher laser energies resonant with the C excitons or unbound electron-hole excitations [21–24]. The electronic excitations at these energies are more likely to be extended over both monolayers and thus more sensitive to interlayer separation distance.

The 2° sample offers another perspective on the moiré physics. Using the measured shifts of the A$_1'$ and its satellite we can obtain a value for the zone-folding wavevector associated with this peak and use it to predict the shifts we would expect for the other phonon branches. The shifts predicted for the TA and LA acoustic branches of the WSe$_2$ layer based on this wavevector are well within the range we would expect to be able to observe; $26.7 \pm 0.6$ and $41.0 \pm 0.9$ cm$^{-1}$. The failure to observe these modes is presumably because they only weakly couple to the intralayer excitons. A similar effect has been observed for the shear mode peak observed in reconstructed WSe$_2$/MoSe$_2$ heterostructures which is a factor of 5-10 weaker for twist angles near 0° compared to twist angles near 60° [22].

Having analysed both the 2 and 57° samples we now have values for $k_c$ for two samples. When we compare $k_c$ with the predictions for the moiré wavevectors [12,25] we surprisingly discover that they are not the moiré wavevector but instead appear to be twice the moiré wavevector. The ratios of the magnitude of the wavevector obtained from the Raman shifts and the moiré wavevector predicted from the measured twist angles are 2.0 (1.7-2.3) for the 57° sample and 2.1 (1.6-2.7) for the 2° sample; the ranges quoted are calculated using the 1° uncertainty in the twist angles. Whilst for the acoustic branch it might be possible to confuse single phonon Raman at twice the moiré wavevector with two phonon Raman involving phonons with the moiré wavevector, this is not possible for the optical phonons. For the 57° sample it is clear that we are not observing an $A_1'$ satellite peak associated with the moiré wavevector, which we would predict would be at 238.8 cm$^{-1}$. It is also clear that for the zone-folded acoustic modes any scattering at the moiré wavevector is much weaker than the scattering at twice the moiré wavevector.

As the observation that zone folding appears to be occurring at twice the moiré wavevector is unexpected, we should consider alternatives to the simple model. One possible alternative is that strain in the TMDC layers has modified the moiré pattern. It has been shown that strains formed in the different layers during the manufacture of heterostructures can be sufficiently large that they are comparable to twist angles of a few degrees and can leading to elongated, 1D, rather than hexagonal, 0D, moiré unit cells and linearly polarised interlayer exciton photoluminescence[26]. The 57° discussed in this paper was included in the study of the effect of strain on PL [13] and was found to one of the ~20% of samples which did not exhibit significant strain effects. In addition, the zone-folding acoustic mode peaks are well defined which would require the strain to be uniform across the sampled area. TEM and scanned probe studies [26–28] indicate that generally the strain is significantly non-uniform even over areas significantly smaller than the 2 μm spots probed in the Raman experiments. Also the strains reported so far have been uniaxial and thus their direction relative to the twist related moiré effects will alter the zone-folding vectors observed. It is difficult to believe that the direction of the strain in two samples will fall fortuitously to give the same correspondence between the measured zone-folding wavevector and twice the expected moiré wavevector predicted assuming the absence of significant strain.

Another complexity of moiré physics is the possibility of reconstruction of the lattice which has been observed in heterobilayers with twist angles near to 0 and 60° [27,28]. Theoretical models of reconstruction have been developed which can predict the range of twist angles for which it would be expected [28,29]. Both samples discussed in this paper are near the range of angles for which significant reconstruction would be expected however they are more likely to be outside this range. If reconstruction was occurring in these samples then the commensurate parts of the lattice would not contribute to moiré effects and the domain walls would be unlikely to have a well-defined moiré periodicity. Therefore, we prefer an explanation which does not rely on reconstruction.

Despite it being unexpected that the zone-folding wavevector is twice the moiré wavevector this is what the experimental results indicate. Theoretically it is possible for this to occur as the excitonic states involved in the Raman scattering have structure on length scales smaller than the moiré periodicity which could suppress scattering by phonons with the moiré wavevector. The fact that similar measurements on $MoS_2$ twisted bilayers can be interpreted using a zone-folding wavevector which is the moiré wavevector adds to the mystery. However it should be noted that the homobilayers experiments were performed on samples with much larger twist angles, between 9 and 49° [12], and so it is not clear if the behaviour reported here is unique to twisted heterobilayers. Further investigations of a range of heterobilayers with different twist angles may allow unique behaviour to be fully understood and may have implications for other small twist angle 2D systems.

In conclusion, these results are the first experimental proof that it is possible to produce superlattice, multiple-periodicity moiré structures using only two van der Waals coupled layers. Whilst current sample production techniques mean that producing one of these structures will require some serendipity it is to be hoped that future improvements in sample production may allow them to be produced to order. In any case we can now to start to explore the possibility of new emergent phenomena in these structures. In addition the possibility of their formation should be considered in the interpretation of the properties of existing samples and characterisation performed to determine if such structuring is present.

These results illustrate the fact that resonant Raman scattering can both quantify the periodicities in these structures and provide information about the strength of coupling of intralayer excitons to the moiré structure. The huge enhancement of the scattering from the zone-folded acoustic modes at the narrow linewidth A excitons suggests that future attempts at observing these modes should use resonant excitation with these states rather than the more common 532 or 633 nm excitation. The resonance behaviour may explain why these modes have not been observed in TMD heterostructures before[21,22].

The observation of well-defined triplet Raman peaks requires that it is possible to optically excite intralayer excitons in the moiré potential with a spatial coherence length of at least the superlattice lattice parameter which for the sample presented here is ~ 9 times the moiré length scale. This indicates that it should be possible to investigate the coherent manipulation of lateral arrays of almost identical excitons. An exciting extension of the work presented here would be to investigate interlayer excitons, possibly weakly hybridised with intralayer excitons, in similar structures[30,31]. These states have much longer lifetimes[32], better coherence and form condensates[33], and therefore could have potential applications in quantum computing. Another exciting possibility would be coherent arrays of interlayer excitons interacting with strongly-correlated electronic states leading to exotic inter-excitonic interactions[3,4,34]. Whilst achieving these goals will require considerable further progress in terms of sample quality and our understanding of the moiré and superlattice physics in these exciting systems they are surely worth the effort.

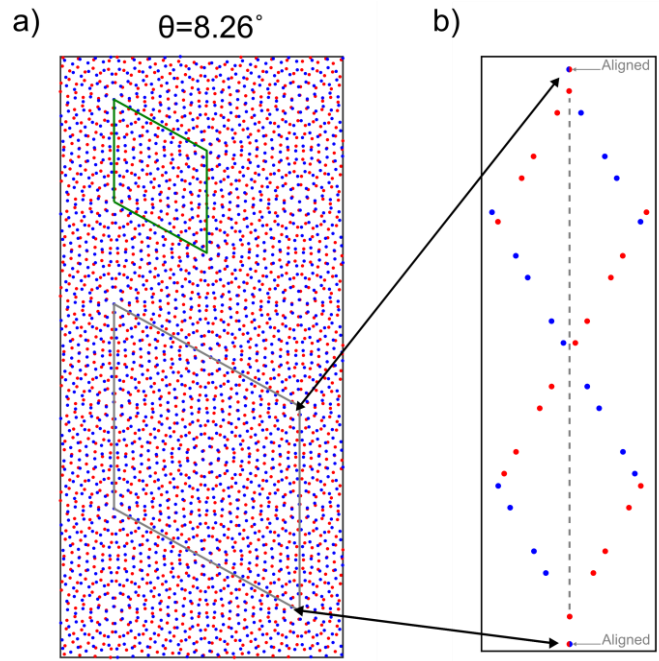

*Figure 1 Illustration of a dual periodicity moiré superlattice a) The moiré interference pattern is shown for a homobilayer with a twist angle of 8.26°, where the atoms in the two layers are indicated by the red and blue points. For this twist angle the moiré period is half that of the superlattice, as demonstrated by the moiré (green) and superlattice (grey) unit cells. **b**) Shows the atomic displacements along the superlattice vector where a translation along this vector results in perfect alignment of the atoms in both layers (indicated by the grey arrows), whereas translation by the moiré lattice vector results in a displacement of the atoms in the different layers. Depending on the sample twist angle dual periodicities with larger ratios of the superlattice and moiré periodicity can arise in twisted bilayer structures.*

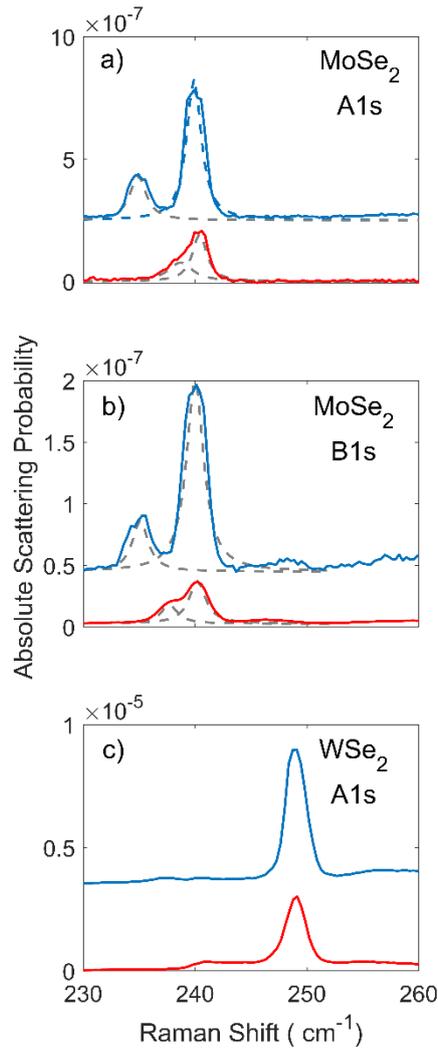

*Figure 2 Raman spectra of Zoned Folded Optical Phonons in 2° and 57° MoSe₂/WSe₂ Heterobilayers* -Raman spectra of two MoSe$_2$/ WSe$_2$ heterobilayers with twist angles of 57 ° (blue) and 2 ° (red) obtained when resonant with the MoSe$_2$ A1s, B1s and WSe$_2$ A1s intralayer excitons; panels (a), (b) and (c) respectively. In the spectra for the 57 ° sample obtained at the MoSe$_2$ resonances we observe the $A_1'(\Gamma)$ phonon at 240 cm$^{-1}$ and a peak at 234.8 cm$^{-1}$. At the same resonances in the 2 ° sample we observe the $A_1'(\Gamma)$ and a new peak at 236.8 cm$^{-1}$ that appears as a shoulder to the stronger peak. Neither of the lower shift peaks are observed in Raman spectra of monolayers of MoSe$_2$. At the WSe$_2$ A1s intralyer exciton resonance the spectra for both heterostructures present only a single strong peak at ~249 cm$^{-1}$ which is due to the $A_1'/E'(\Gamma)$ phonon.

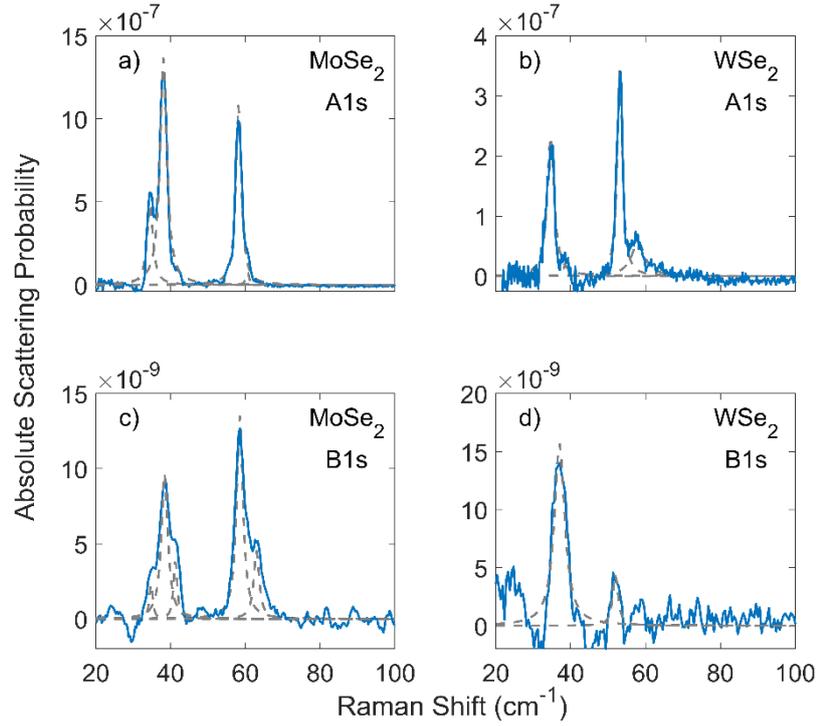

*Figure 3 Raman spectra of Zoned Folded Acoustic Phonons in a 57° MoSe$_2$/WSe$_2$ Heterobilayer -* Low frequency Raman spectra obtained from the 57 ° sample when resonant with the MoSe$_2$ and WSe$_2$ A1s and B1s intralayer excitons. At the MoSe$_2$ A1s (a) and B1s (c) resonances we observe at least 5 Raman peaks in two clusters in the ranges 34 to 42 cm$^{-1}$ and 56 to 65 cm$^{-1}$. Likewise, at the WSe$_2$ A1s (b) we observe at least 4 peaks appearing in two clusters with the dominant peaks at 34.6 and 59.2 cm$^{-1}$ along with two weaker peaks. These peaks appear as higher frequency shoulders on the dominant peaks. At the WSe$_2$ B1s (d) resonance we observe a weak, broader feature near 36-40 cm$^{-1}$ which can be explained using the lower pair of peaks observed at the WSe$_2$ A1s resonance. It should be noted that the WSe$_2$ B1s intralayer exciton in this sample demonstrates an ~ 10% hybridisation with an interlayer exciton.

|  | Peak No. | TA | | LA | | A$_1$ | |
| --- | --- | --- | --- | --- | --- | --- | --- |
|  |  | Exp | Theory | Exp | Theory | Exp | Theory |
| **WSe$_2$** | K$_{S-}$ | - | 31.4±0.4 | - | 47.9±0.5 | - | 6.8±0.1 |
|  | K$_C$ | 34.6±0.1 | 34.9±0.1 | 52.9±0.1 | 52.9±0.1 | - | 7.5±0.1 |
|  | K$_{S+}$ | 38.8±1.1 | 38.4±0.3 | 57.8±0.5 | 57.8±0.5 | - | 8.1±0.1 |
| **MoSe$_2$** | K$_{S-}$ | 35.4±0.2 | 34.1±0.4 | - | 53.5±0.6 | - | -4.1±0.1 |
|  | K$_C$ | 39.0±0.2 | 37.9±0.1 | 59.2±0.1 | 59.2±0.1 | -5.1±0.1 | -5.1±0.1 |
|  | K$_{S+}$ | 42.3±0.1 | 41.7±0.4 | 63.7±0.5 | 64.8±0.6 | - | -6.2±0.1 |

*Table 1* Presents the experimental and predicted frequencies of both the acoustic and optical zone folded phonon observed in the 57 ° heterobilayer. Each peak has been assigned to its appropriate underlying phonon branch (TA, LA, A$_1$'). Predicted frequencies were obtained using the wavevectors extracted from the phonon dispersion relations for the WSe$_2$ peaks at 52.9 and 57.8 cm$^{-1}$ shown in red. We associate Kc with the dominant peak at 52.9 cm$^{-1}$ and then determine the wavevectors of the two sidebands (K$_{S+}$ and K$_{S-}$) using the difference between our two wavevectors. Experimental errors shown are a standard deviation determined from fitting multiple spectra. Errors shown on predicted values were obtained by propagating the experimental errors in the frequency of the 52.9 and 57.8 cm$^{-1}$ peaks.

**Methods**

The samples used consist of mechanically exfoliated monolayers of $MoSe_2$ and $WSe_2$ stacked to form Van der Waals heterobilayers and encapsulated between layers of hexagonal boron nitride using a dry transfer technique [35] with the underlying substrate consisting of oxide coated silicon. For low temperature measurements the samples were mounted inside an Oxford Instruments High Resolution liquid helium flow microstat, and unless stated otherwise all experiments were performed at 4 K. All optical measurements were carried out using a back-scattering geometry with a x50 Olympus objective producing a spot size on the sample of 3 µm. Positioning on the sample was achieve a computer controlled 3 axis translation stage and a custom in situ microscope. For all measurements the incident power on the sample was maintained below 100 µW to avoid photo doping and laser heating of the sample [36,37]. The resonance Raman measurements were carried out using a CW Coherent Mira 900 allowing excitation energies from 1.24 to 1.77 eV and a Coherent Cr 599 dye laser using DCM, Rhodamine 6G and Rhodamine 110 laser dyes allowing excitation energies from 1.74 to 2.25 eV. In addition, for PL measurements a 532 nm Coherent Verdi laser was also utilized. The incident polarizations of the lasers were horizontal relative to the optical bench and the Raman scattered light coupled into the spectrometer was analyzed using both horizontal and vertical polarizations. The polarization of the Raman peaks was observed to be strongly co-linear. This allowed unwanted luminescence from the samples to be removed from the Raman spectra by subtraction of the crossed and parallel polarized spectra. Both Pl and Raman spectra were measured using a Princeton Instruments Tri-vista Triple spectrometer equipped with a liquid nitrogen cooled CCD. The Raman peak frequencies were all calibrated using the silicon Raman peak at 520 $cm^{-1}$ as an internal reference. To allow comparison of the Raman scattering on the $MoSe_2$ and $WSe_2$ samples the spectra were calibrated to absolute Raman scattering probability. This required the normalization of the Raman spectra to the 520 $cm^{-1}$ Silicon peak intensity, correction of Fabry-Perot interference effects and calibration using the absolute Raman scattering results of Aggarwal et al [16]. To account for the Fabry-Perot interference we made use of reflectivity spectra measured using a Fianium super continuum source and Ocean optics HR4000 spectrometer.

## Supporting Information

The supplementary information contains: a discussion of background subtraction methods; additional Raman spectra for heterobilayers when resonant with the $MoSe_2$ and $WSe_2$ intralayer excitons; DFT predictions of the phonon dispersion relations; details of the predicted moiré wavevector; a discussion of additional zone folded optical phonons and analysis of the resonance Raman behavior of the zone folded Raman peaks.

## Data Availability

The data presented in this paper is openly available from the University of Southampton Repository DOI: https://doi.org/10.5258/SOTON/D1546

## Author Contributions

The experiments were conceived by D.C.S, L.P.M and X.X. Samples were fabricated by P.R. The experimental measurements were performed by J.V and L.P.M. Experimental data analysis and interpretation was carried out by L.P.M, D.C.S and J.V. DFT predictions of heterobilayers and monolayer phonon dispersion relations was calculated by H.S and A.C. The paper was written by D.C.S and L.P.M. All authors discussed the results and commented on the manuscript.

## Corresponding Author

*D.C.Smith@soton.ac.uk

## Competing financial interests

The authors declare no competing financial interests.

**Funding Sources**

Research at the University of Southampton was supported by the Engineering and Physical Science Council of the UK via program grant EP/N035437/1. Both L.P.M and J.V were also supported by EPSRC DTP funding. The work at University of Washington was mainly supported by the Department of Energy, Basic Energy Sciences, Materials Sciences and Engineering Division (DE-SC0018171). The work at Czech Technical University in Prague was done with the support of the project "Novel nanostructures for engineering applications" No. CZ.02.1.01/0.0/0.0/16_026/0008396, and by The Ministry of Education, Youth and Sports from the Large Infrastructures for Research, Experimental Development and Innovations project "e-Infrastructure CZ - LM2018140".